\documentclass[a4paper,twoside,12pt]{article}

\usepackage{graphicx}
\usepackage{psfig}
\usepackage{amsfonts}
\usepackage{amssymb}

\title{Dynamical vacuum selection in field theories with flat
directions in their potential}
\author{J.Striet$^1$ and F.A. Bais$^2$\\[2mm]
Institute for Theoretical Physics \\ University of Amsterdam \\
Valckenierstraat 65 \\ 1018XE Amsterdam \\ The Netherlands\\\date{November 27, 2002}}

\begin{document}
\maketitle
\footnotetext[1]{jelpers@science.uva.nl}
\footnotetext[2]{bais@science.uva.nl}


\begin{abstract}
\noindent In this paper we show that in field theories with
topologically stable kinks and flat directions in their potential, a
so-called dynamical vacuum selection (DVS) takes place in the
non-trivial, soliton sector of the theory. We explore this DVS
mechanism using a specific model. For this model we show that there is
only a static kink solution when very specific boundary conditions are
met, very similar to the case of vortices in two dimensions. In the
case of other boundary conditions a scalar cloud is expelled to
infinity, leaving a static kink behind. Other circumstances under
which DVS may or may not take place are discussed as well.
\end{abstract}

\section{Introduction}
In this paper we examine topological defects in theories with flat
directions in their potential. Flat directions in the scalar potential
are quite natural in the context of supersymmetric models. As was
noticed in \cite{witten2}, in spite of the fact that a model does
allow topologically stable vortices, not all admissible boundary
conditions in a given model with flat directions, are compatible with
the existence of a static vortex solution. In \cite{penin} such
vortices in theories with flat directions were studied, and it was shown
that in the presence of a topologically stable vortex, a specific
vacuum is dynamically selected. In this paper we focus on the one
dimensional case and show that also for theories in one dimension
with flat directions not all boundary conditions allow the existence
of a static kink. We will show that a specific vacuum is dynamically
selected in the presence of such a kink or domain wall.  Although we
will use a specific model, it will become clear that dynamical
vacuum selection (DVS) is a general feature for theories with
flat directions in one dimension.\\[2mm] To be a bit more specific we
will discuss a class of models which have two copies of a scalar Higgs
field and a potential of the following form, 
\begin{equation}
 V(\phi_1,\phi_2) = \frac{\lambda}{4}
(\phi_1^2-\phi_2^2-f^2)^2\quad, 
\end{equation}
in analogy with the model studied in \cite{penin}.

In section \ref{1D} we focus on the one dimensional model, in section
\ref{2D} we comment on the two dimensional model which was discussed
in \cite{penin}, and in section \ref{3D}, we show with the help of a
Bogomolny \cite{bogo} type argument that there is no DVS in the three
dimensional case, as was already anticipated in \cite{penin}. We end
the paper with some conclusions and a brief discussion.

\section{Dynamical vacuum selection (DVS) in 1 dimension}
\label{1D}
In this section we investigate DVS in a specific one dimensional
model. First we introduce the model, subsequently we prove that there
is only a static kink solution for a very specific non trivial
boundary condition out of a continuum of allowed boundary
conditions. Finally we investigate the kink dynamics if this specific
boundary condition is not met and we find that indeed the DVS
mechanism becomes operative.

\subsection{The model}
We consider a model with two real scalar fields and a potential which
allows for the formation of topologically stable kinks and which furthermore
features a flat direction.\\ The model is given by:
\begin{equation}
\mathcal{L} = \int dx \{\frac{1}{2}(\partial_\mu\phi_1)^2 +
\frac{1}{2}(\partial_\mu\phi_2)^2 - 
\frac{\lambda}{4}(\phi_1^2-\phi_2^2-f^2)^2 \}\quad,
\end{equation}
with $\phi_1$ and $\phi_2$ two real scalar fields.

It is quite clear that this model contains topologically stable kinks,
which have to satisfy the spatial boundary conditions
$\phi_1(\pm\infty)=\sqrt{f^2+\phi_2(\pm\infty)^2}$ and
$\phi_1(\mp\infty)=-\sqrt{f^2+\phi_2(\mp\infty)^2}$. Note that the
boundary values of $\phi_2$ do not influence the topological charge of
the kink. Thus there is a two parameter class of topologically stable
kinks present in this model, labeled by
$(\phi_2(-\infty);\phi_2(+\infty))$. In the next section we
investigate the class of static kink solutions in this model when real
space is taken infinite, this class as we will show is in fact very
small.

\subsection{Static kinks}
\label{static}
Naively one might expect to be able to find a static kink solution
(assuming real space to be infinite) for any set of the boundary values
of $\phi_2$ in the topologically nontrivial sector. However this turns
out not to be true. What we in fact will show is that there is only
one very specific set of boundary values of $\phi_2$ for which a
static kink solution exists.

To obtain a static configuration, one sets the time derivatives equal
to zero and extremizes the resulting Hamiltonian. This Hamiltonian of
this model can be interpreted as the action of a point particle in a
two dimensional potential, and we may analyze the system through this
mechanical analogue. More explicitly, after making the following
identifications: $x\to t$, $\phi_1\to x$ and $\phi_2\to y$, we get the
following action:
\begin{equation}
S=\int dt \{ \frac{1}{2}(\partial_tx)^2 + \frac{1}{2}(\partial_ty)^2 +
\frac{\lambda}{4}(x^2-y^2-f^2)^2 \}\quad.
\end{equation}
This system corresponds to a point particle moving in the potential
$-\frac{\lambda}{4}(x^2-y^2-f^2)^2$.

\begin{figure}[!htb]
\begin{center}
\includegraphics[width=8cm,angle=270,clip]{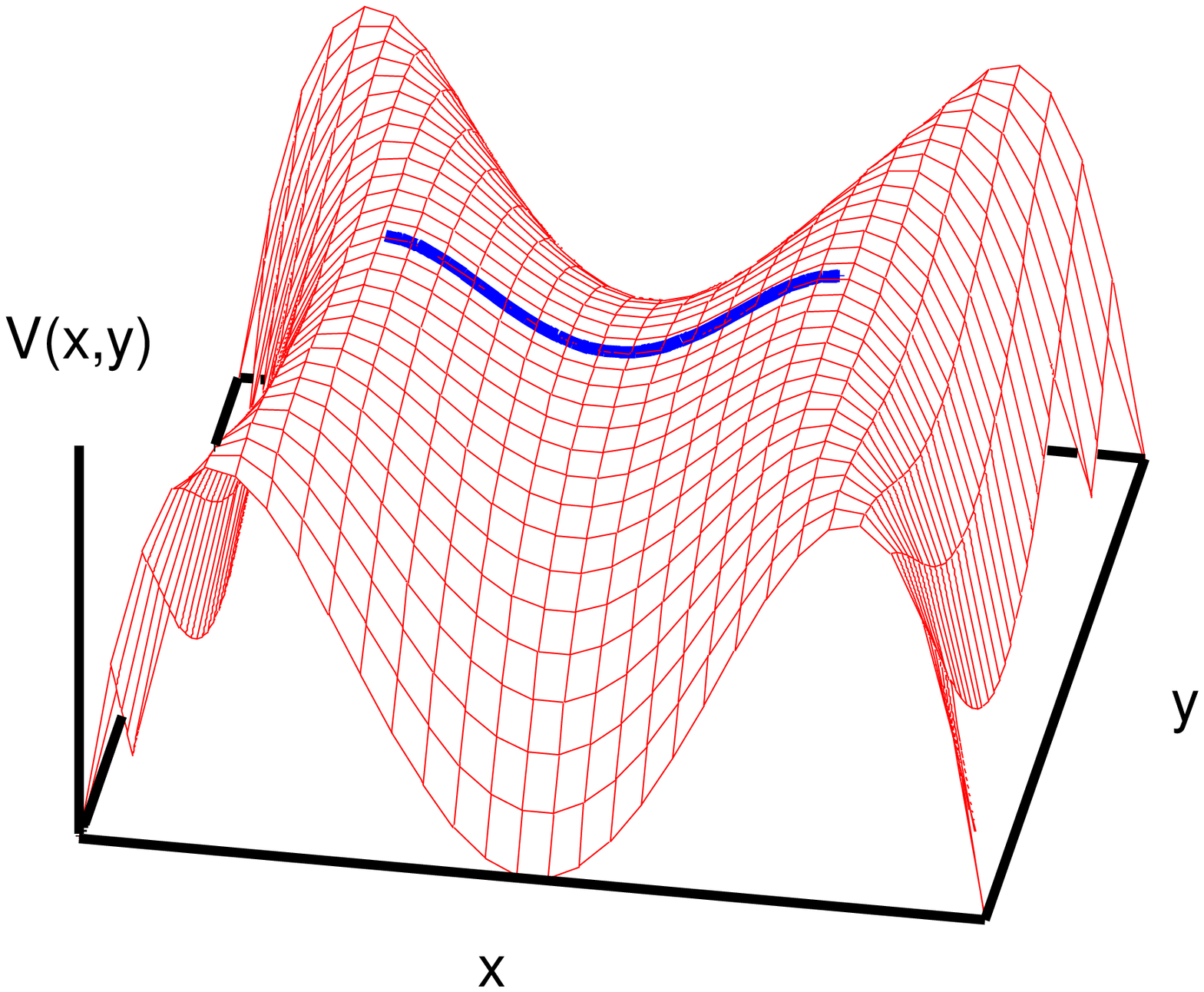}
\caption[somethingelse]{\footnotesize The two dimensional potential
for the point particle. The only path satisfying the desired boundary
conditions is the path with $y(t) =0$. This path represents the only
static kink solution in an infinite space, for which $\phi_2(x)=0$.}
\label{potential.eps}
\end{center}
\end{figure}

The problem of finding a static kink solution is now translated to
finding a solution to the equations of motion of the point particle,
which moves from one point on one line of maxima of the potential,
$x=\pm\sqrt{f^2+y^2}$, at $t\to-\infty$ to an other point on the other
line of maxima, $x=\mp\sqrt{f^2+y^2}$, at $t\to \infty$ (see figure
\ref{potential.eps}).

At $t\to-\infty$ the particle should be at rest in one of the maxima
of the potential. At $t\to\infty$ the particle needs to be at rest in
one of the other set of maxima of the potential. Obviously there is
only one path which satisfies these these boundary conditions and that
is the path with $y=0$. Any path starting with $y(t\to-\infty)\neq0$
in one of the maxima of the potential will never again reach a maximum of the
potential. In such a case it is easy to prove that the kinetic
energy of the particle associated with component of the motion in the
$y$-direction will increase monotonically in time and therefore the
particle can never climb out of the potential well again, i.e., will
not not able to satisfy the boundary condition at $t\to\infty$.
Translating this back to the kink solutions, this means that one can
only have a static kink solution if the boundary conditions of
$\phi_2$ are $\phi_2(\pm\infty)=0$ and moreover for the static
solution we get $\phi_2(x)=0$. Thus the only static kink solution for
this model in an infinite space is equal to the usual kink solution.

A crucial step in deriving this result was the infinite size of real
space. If space is finite the argument changes dramatically. In the
mechanical analogue this would mean that the particle can have an
initial velocity (and direction of this velocity), so that the entire
class of boundary configurations is allowed.

A natural question to ask is, what happens if the boundary conditions
of $\phi_2$ are {\em not} of the specific form
$\phi_2(\pm\infty)=0$. As we just demonstrated, there can not be a
static kink solution with these boundary conditions and we are led to
ask how the configuration will develop in time. Later on we will study
the dynamics of such a configuration and find that DVS will take
place. Before we turn to this question we take a closer look at the
structure of the static kink solutions in a finite space and at how
the boundary conditions effect the core structure of the kink.

\subsection{Kinks in finite space.}
In this section we study static kink solutions in a finite
space with fixed boundary conditions. These kinks correspond to the so
called restricted instantons in quantum mechanics \cite{affleck}.
We first want to introduce the massless
modulus field, $z$, which is of paramount interest
to us in the rest of the paper. In the broken phase of the theory this
field corresponds to the degree of freedom in the flat direction of
the potential. On the vacuum manifold we have
$\phi_1^2-\phi_2^2=f^2$. We can parameterize the degree of freedom in
the flat direction of the potential by writing: $\phi_1=f\cosh{u}$ and
$\phi_2=f\sinh{u}$. To get the canonical kinetic term we change from
$u$ to the modulus field $z$, which is given by:
\begin{equation}
z=f \int_0^udu^\prime\sqrt{\cosh{2u^\prime}}\quad.
\end{equation}
This field $z$, obeys the free massless equations of motion. From the
construction for this specific case it is evident that there is always
a massless mode if there is a flat direction in the potential. The
dynamics and statics (energy) of this massless mode are the crucial
ingredients in DVS. They allow one to show directly, that there can be
DVS in one and two dimensions but not in three or higher dimensions.

Let us return to the restricted kinks. It is not hard to anticipate
what the solution of a kink will look like if the size of the space,
$2 R_\infty$, is much larger than the core size of the kink, $2
R_c$. Consider the configuration of a kink of size $2 R_c$, where
outside the core of the kink the vacuum manifold is reached
exponentially fast. To this kink we add a tail at each side in which
the scalar fields stay in the vacuum manifold but move to the
prescribed boundary value at $x=\pm R_\infty$. In the limit of
$R_\infty\to\infty$ we should recover the unique static solution we
found before. Thus the kink solution we should use to describe the
core of the kink is this special case where $\phi_2=0$.

\begin{figure}[!htb]
\begin{center}
\includegraphics[width=6cm,angle=270,clip]{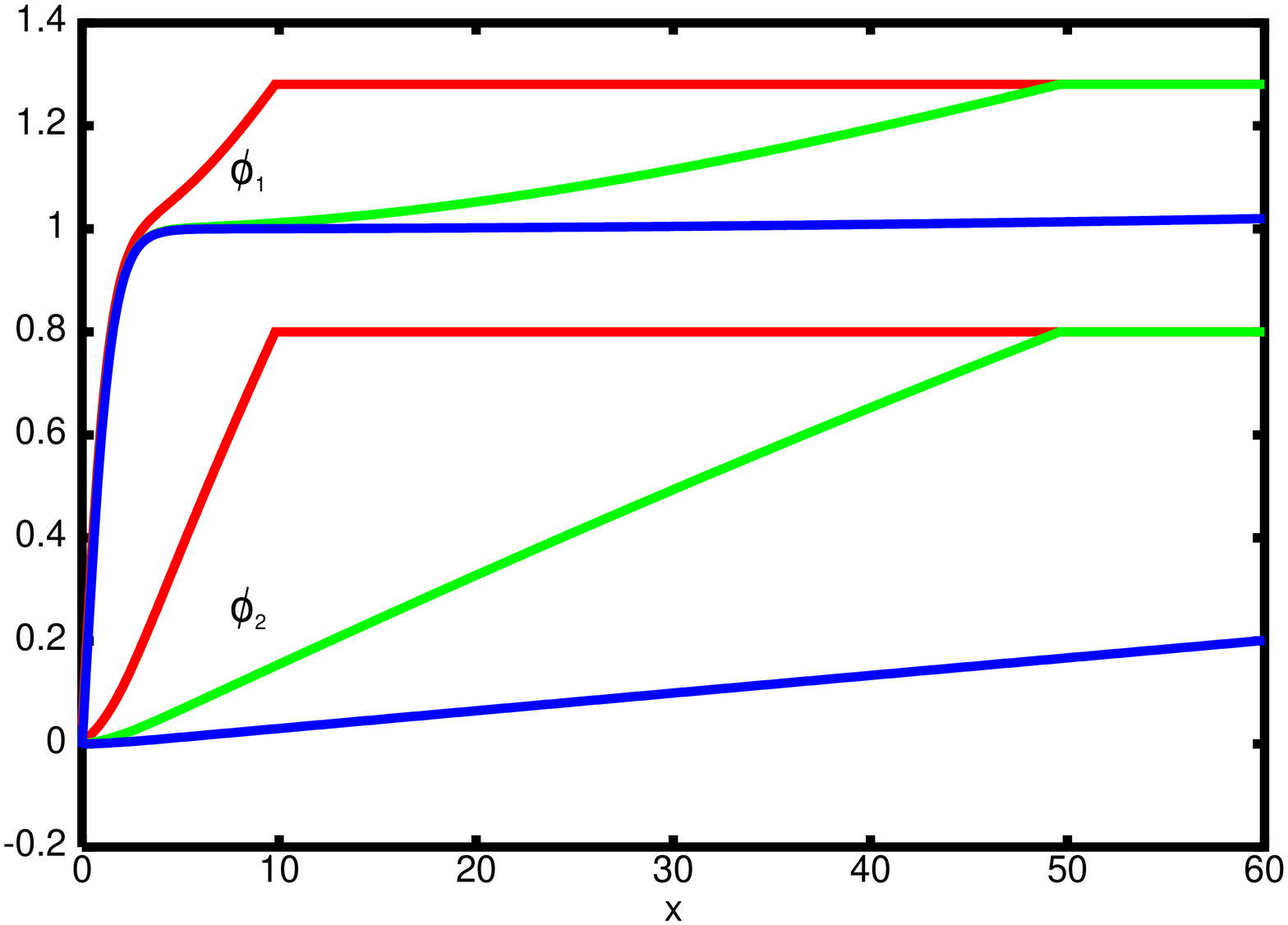}
\caption[somethingelse]{\footnotesize This figure shows the fields for
the restricted kinks for different values of $R_\infty$. The values of
$R_\infty$ are $R_\infty=10$, $R_\infty=50$ and $R_\infty=250$. The
boundary conditions on $\phi_2$ are $\phi_2(\pm R_\infty)=\pm
0.8$. The fields for negative values of $x$ just follow from symmetry
of the configuration and are not plotted. This figure clearly shows
the separation of the solution in the kink and a tail in the limit of
small $z_\pm/R_\infty$.}
\label{restrictedkinks.eps}
\end{center}
\end{figure}

So we approximate the restricted kink solution by a configuration
which is a superposition of two linear tails and a kink with
$\phi_2=0$. Clearly the tails and the kink independently satisfy the
field equations, so it is only due to the overlap that the field
equations are not quite satisfied. The violation to the equations of
motion due to the overlap, is proportional to $z_\pm/R_\infty$ in the
action, and this justifies the approximation we made in the limit of
small $z_\pm/R_\infty$. With the help of a relaxation program we
numerically determined the static kink solution for various values of
the parameters, see figure
\ref{restrictedkinks.eps}. In all our figures and numerical
simulations we take $\lambda=f=1$, any other values of $\lambda$ and
$f$ follow by rescaling the fields and the space coordinate.

Using these approximate restricted kink solutions we can also
determine the position of the kink with respect to the boundaries of
the space. We can get an estimate by minimizing the energy in the
tails. Both tails want to spread as much as they can to lower the
energy, so if there is an asymmetry in the boundary conditions of
$\phi_2$, this will certainly have an effect. We approximate the
position of the kink by minimizing the sum of the energy of the two
tails, which is given by:
\begin{equation}
E_{tail} = \frac{z_+^2}{R_+} + \frac{z_-^2}{R_-}\quad,
\end{equation}
with $R_+ + R_-=2R_\infty$ and $z_\pm$ the values of the modulus field
at the boundaries of space.

\begin{figure}[!htb]
\mbox{\psfig{figure=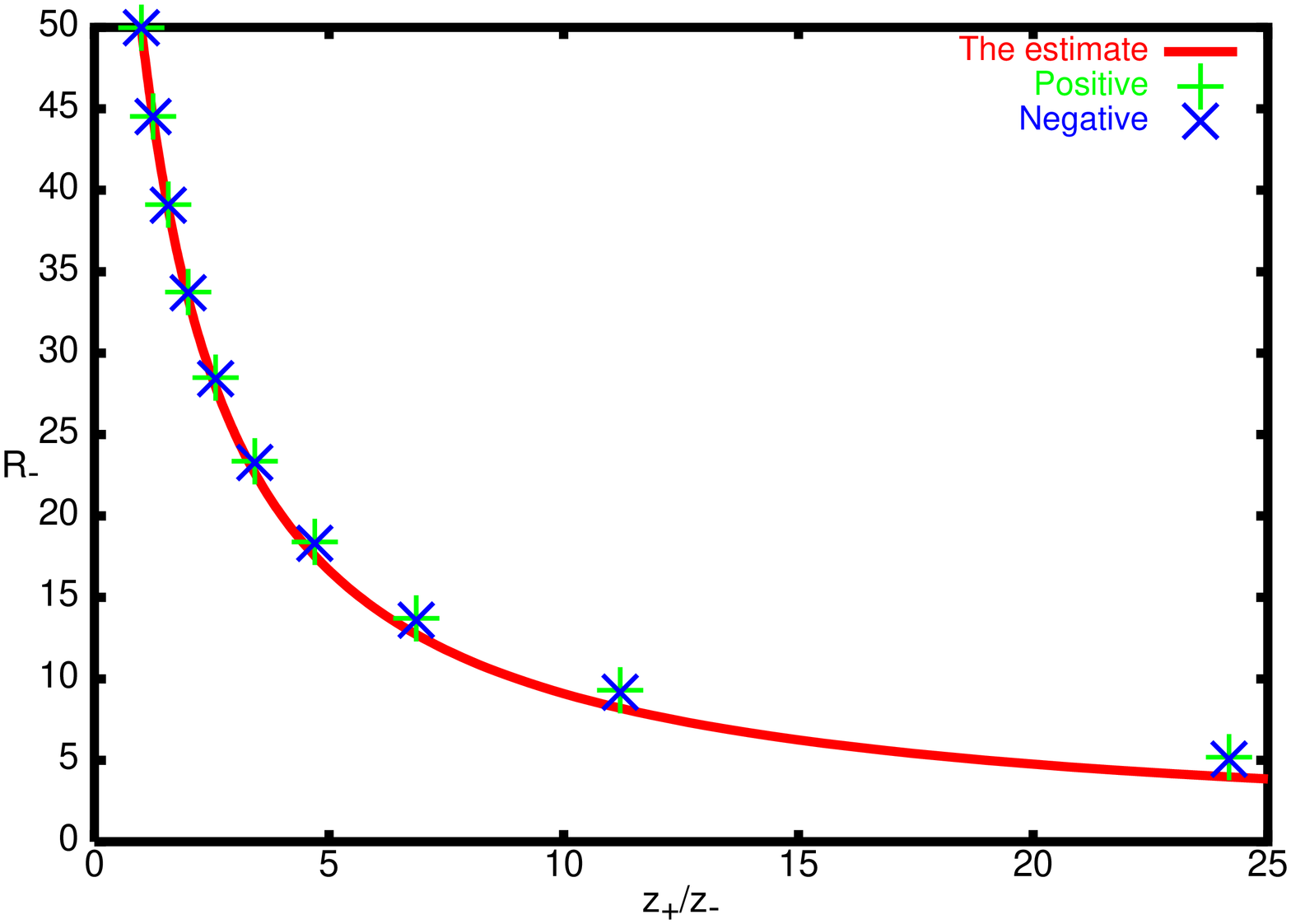,angle=270,width=7.9cm}}
\mbox{\psfig{figure=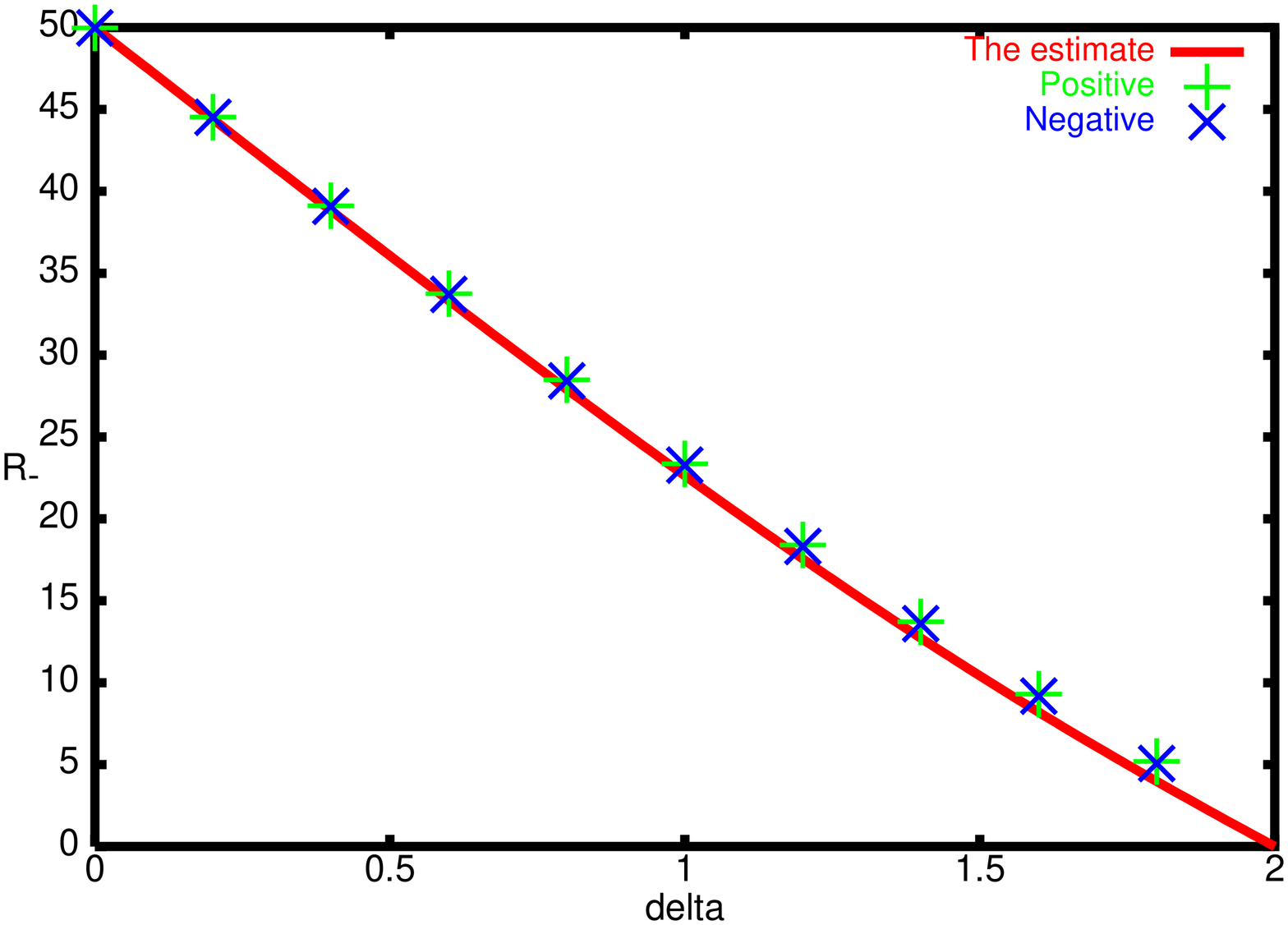,angle=270,width=7.9cm}}
\makebox[7.9cm][c]{\footnotesize{(a)}}
\makebox[7.9cm][c]{\footnotesize{(b)}}
\caption[somethingelse] 
{\footnotesize \\ The figures (a) and (b) show the estimate for $R_-$
and the numerically obtained values of $R_-$ as a function of
$z_+/z_-$, figure (a), and of $\delta$, figure (b), with
$R_\infty=50$. The boundary values of $\phi_2$ are
$\phi_2(R_\infty)=2+\delta$ and $\phi_2(-R_\infty)=\pm
2\mp\delta$. The figures show that the estimate of the position of the
kink works very well and that the position of the kink is independent
of the sign of $\phi_2(-R_\infty)/\phi_2(+R_\infty)$ as prescribed by
our estimate.}
\label{positionofkinkz.eps}
\label{positionofkink.eps}
\end{figure}

Minimizing this energy under the restriction $R_+ + R_- = 2R_\infty$
gives:
\begin{equation}
R_+ = \frac{2}{1+|\frac{z_-}{z_+}|}R_\infty \quad;\quad R_- =
\frac{2}{1+|\frac{z_+}{z_-}|}R_\infty \quad,
\end{equation}
with $R_+$ and $R_-$ the distance between the core of the kink and the
$+R_\infty$ and $-R_\infty$ boundary of space respectively. Note that
the position of the kink in this estimate does not depend on the
relative sign between the boundary conditions on $\phi_2$ at
$+R_\infty$ and $-R_\infty$. We tested this simple estimate
numerically and found it to work quite well, see figure
\ref{positionofkink.eps}. It should be clear that for $R_\pm<R_c$ the
estimates of $R_+$ and $R_-$ break down.

In figures \ref{positionofkinkz.eps}(a) and
\ref{positionofkink.eps}(b) we plotted the numerical data for the
position of the kink and the estimated position of the kink. We show
the results of numerical simulations for $R_\infty=50$, where we
defined the position of the kink by the zero of the $\phi_1$
field. The boundary conditions we put on $\phi_2$ are
$\phi_2(R_\infty)=2+\delta$ and $\phi_2(-R_\infty)=\pm 2\mp\delta$,
with $\delta$ running from zero to two. In figure
\ref{positionofkinkz.eps}(a) we plotted $R_-$ as a function of
$z_+/z_-$, where in \ref{positionofkink.eps}(b) we plotted $R_-$ as a
function of $\delta$. The plots show a good agreement between the
estimate and the numerical data we obtained. They also show the
independence of the position of the kink on the sign of $z_-/z_+$,
which is equal to the sign of $\phi_2(-R_\infty)/\phi_2(+R_\infty)$.

In the limit of small $z_\pm/R_\infty$ we have a good understanding of
the static kink solution. In the next section we will study the
dynamics of the non static kink solutions, which occur if the boundary
values of the $\phi_2$ field are not zero.

\subsection{Non static kink configurations}
In section \ref{static} we proved that in an infinite space there is
only a static kink solution to the field equations for a specific set
of boundary values of $\phi_2$, notably $\phi_2(\pm\infty)=0$. In any
other case there is no static kink solution to the equations of
motion. In this section we will investigate the behavior of these non
static kinks and find the DVS.\\[2mm] The modulus field $z$ obeys the
free massless equation of motion, which in one dimension is given by:
\begin{equation}
-\partial_t^2 z(x,t) + \partial_x^2 z(x,t) = 0 \quad.
\end{equation}
Thus the modulus field is given by $z(x,t)=z_r(t-x) + z_l(t+x)$. This
shows that the modulus field propagates with the speed of light. Next
we will look at some specific dynamical simulations. In the light of
the previous discussion of constrained kinks it is clear what should
happen in an infinite space. The kink will 'eject' a so called
scalar cloud, the tail, to infinity; the cloud will move with the
speed of light, and will dynamically select the vacuum with
$\phi_2=0$.

We look at two types of initial configurations. One corresponds with the
solution of a restricted kink, which we
numerically determined in the previous section. The other initial
configuration is a configuration with constant $\phi_2$ and for
$\phi_1$ the static kink solution with $f^2$ replaced by
$f^2+\phi_2^2$, see equation (\ref{initial}). Both types of initial
configurations we assume to be at rest at $t=0$.

\begin{figure}[!htb]
\mbox{\psfig{figure=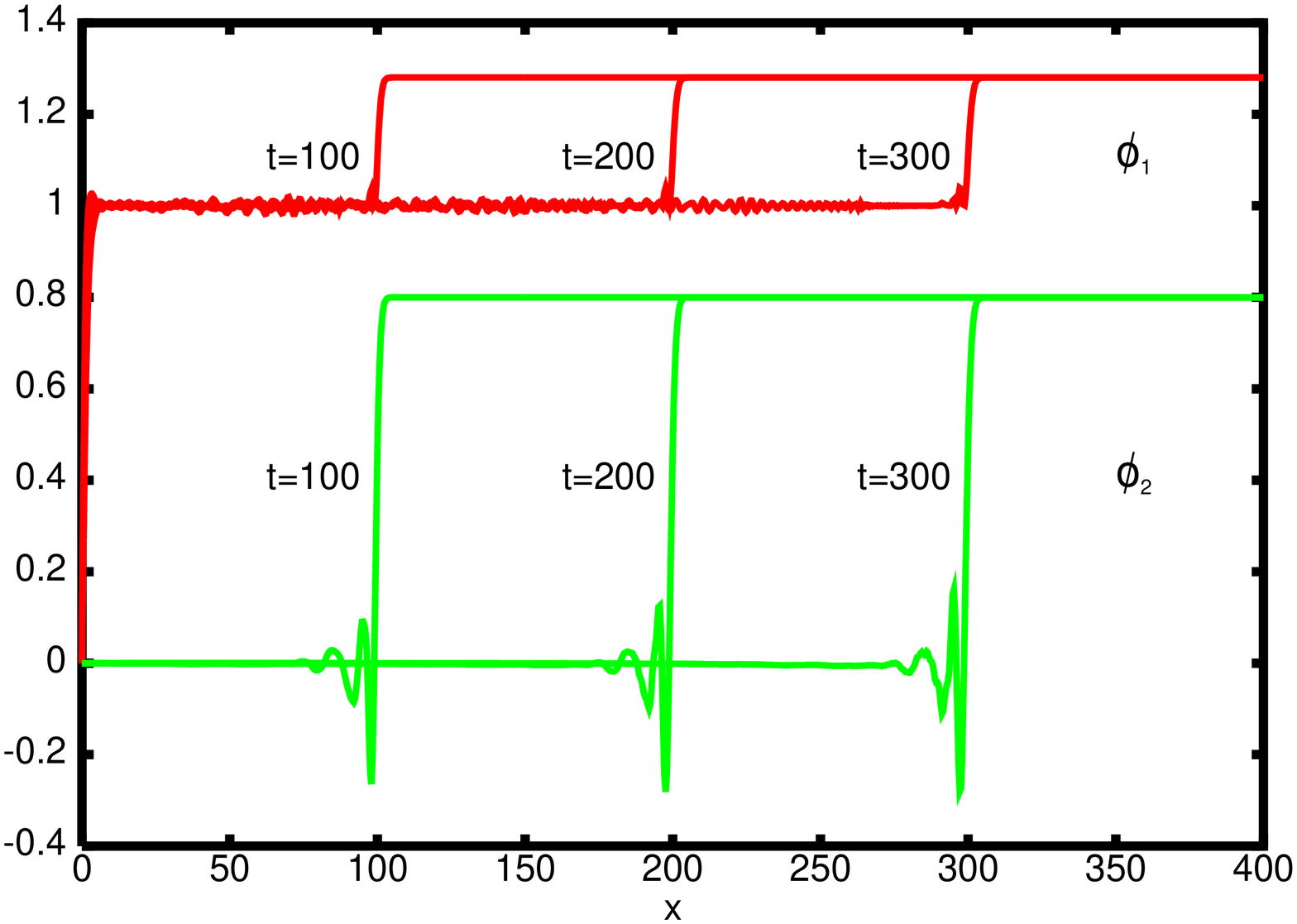,angle=270,width=7.9cm}}
\mbox{\psfig{figure=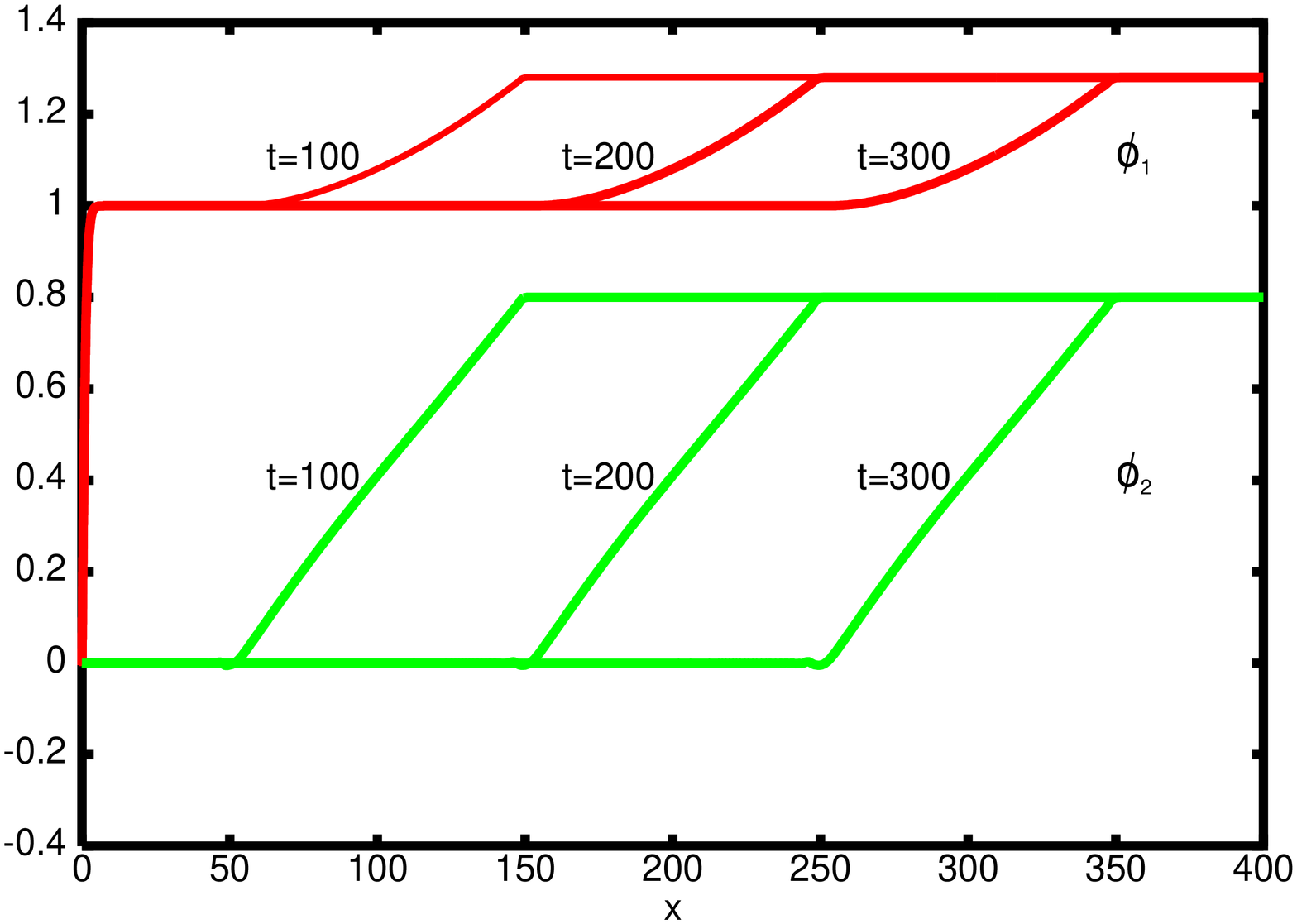,angle=270,width=7.9cm}}
\makebox[7.9cm][c]{\footnotesize{(a)}}
\makebox[7.9cm][c]{\footnotesize{(b)}}
\caption[somethingelse] 
{\footnotesize \\ (a):The dynamics of the kink with
$\phi_2(x)=\phi_2=0.8$ and $\phi_1(x)$ given by equation
(\ref{initial}) at $t=0$. The figure shows snapshots of the fields
$\phi_1(x)$ and $\phi_2(x)$ at $t=100$, $t=200$ and $t=300$. This
shows the DVS and the speed of the scalar cloud.\\ (b):The dynamics of
the restricted kink with $\phi_2(\pm R_\infty)=0.8$ and $R_\infty=50$
at $t=0$. The figure shows snapshots of the fields $\phi_1(x)$ and
$\phi_2(x)$ at $t=100$, $t=200$ and $t=300$. This shows the DVS and
the speed of the scalar cloud.\\ The values of the fields for negative
values of $x$ follow from symmetry of the configuration and are not
plotted.}
\label{dvsofkink.eps}
\label{dvsofreskink.eps}
\end{figure}

Lets us first examine the latter case. Again in our numerical
simulations we take $\lambda=f=1$, and start with:
\begin{equation}
\phi_2(x)=\phi_2\quad;\quad \phi_1(x)=\sqrt{1+\phi_2^2}
\tanh(\frac{1}{2}\sqrt{2}\sqrt{1+\phi_2^2}~x),
\label{initial}
\end{equation}
taking $\partial_t \phi_1(t)|_{t=0}=\partial_t
\phi_2(t)|_{t=0}=0$. As expected, we find that the scalar cloud moves
away with the speed of light and the special vacuum with $\phi_2=0$ is
dynamically selected, see figure \ref{dvsofkink.eps}(a).

The initial condition with the restricted kink as the initial
configuration yields a similar result. We determined the solution of
the restricted kink, with $R_\infty=50$ and $\phi_2(\pm R_\infty)=0.8$,
with the help of a relaxation program and used it as the initial
configuration of the dynamical process. Again we take
$\partial_t\phi_2(t)|_{t=0}=\partial_t\phi_1(t)|_{t=0}=0$. Also here
the special vacuum with $\phi_2=0$ is dynamically selected, see figure
\ref{dvsofreskink.eps}(b).

After the vacuum has been selected the kink can still be excited, which
is clear in the first case, equation (\ref{initial}). In the case
where the restricted kink has been chosen as initial configuration, 
the kink is only slightly excited locally, due to the overlap of the kink and
the modulus field. We conclude that this part of the dynamics
depends on the initial condition, but does not effect the vacuum
selection part of the dynamics.

\section{DVS in 2 dimensions}
\label{2D}
In this section we briefly recall known results of DVS in two
dimensions, as discussed in \cite{penin}. Witten in \cite{witten2}
already observed, that in several models not all boundary conditions
allow for static vortex solutions. In the model, studied in
\cite{penin} there is also a potential of the form
$\frac{\lambda}{4}(|\phi_1|^2-|\phi_2|^2-f^2)^2$. The Higgs fields
they use are now complex scalar fields, oppositely charged under a
local $U(1)$. Also in this model the vacuum $\phi_2=0$ is dynamically
selected in a topologically nontrivial sector of the theory. In
\cite{achucarro} it was pointed out that this specific vacuum was
selected as to minimize the mass of the massive gauge boson.

Again the main idea behind the DVS is the fact that one can make a
tail with the modulus field, which brings fields from one point of the
vacuum manifold to an other point in the same connected component of
the vacuum manifold. In two dimensions the energy cost of such a tail
is inversely proportional to $\ln{\frac{R_\infty}{R_c}}$ and can be
made arbitrarily small in an infinite space. Although the dynamics of
the modulus field is a bit different in two dimensions from
the dynamics in one dimension, the conclusion is the same: DVS takes
place. For more details on DVS in two dimensions we refer to
\cite{witten2}, \cite{penin} and \cite{achucarro}.

\section{No DVS in 3 dimensions}
\label{3D}
To conclude we look at a specific model in three dimensions to argue
that DVS does not work in three dimensions (as was already mentioned
in \cite{penin}). The crucial observation here is that a tail in which
the modulus field connects one point of the vacuum manifold to another
point in the same connected component, will cost a finite amount of
energy. In three dimensions the energy of such a modulus field is
proportional to $\frac{R_cR_\infty}{R_\infty-R_c}$, which does not go
to zero in the limit of $R_\infty\to\infty$, suggesting that there is
no DVS in three dimensions.

To make this more explicit we discuss a model which has a local
$SU(2)$ symmetry and two scalar Higgs fields in the vector
representation of the local gauge group. The potential will again have
the form $\frac{\lambda}{4}(Tr\{\phi_1^2\}-Tr\{\phi_2^2\}-f^2)^2$. In
this model the $SU(2)$ gauge symmetry is spontaneously broken to
$U(1)$ and topologically stable magnetic monopoles can form. The
action of the model is given by:
\begin{equation}
S=\int d^4x \{ \frac{1}{4} F_{\mu\nu}^a F^{a,\mu\nu} +
\frac{1}{2}Tr\{(D_\mu\phi_1)^2\}+ \frac{1}{2}Tr\{(D_\mu\phi_2)^2\} -
\frac{\lambda}{4}(Tr\{\phi_1^2\}-Tr\{\phi_2^2\}-f^2)^2 \}\quad.
\end{equation}
We will try to find static monopole solutions in the BPS limit, i.e.,
we keep the boundary terms fixed but put $\lambda$ to zero. We note
that the fields $\phi_1$ and $\phi_2$ need to be parallel to each
other in the internal space at spatial infinity in order to have a
finite energy solution, and to have an unbroken $U(1)$ in the first
place. We will focus on static configurations where
$Tr\{\phi_2(r\to\infty)^2\}$ does not depend on the spatial angles.

To find static solutions we have to extremize the energy:
\begin{equation}
E=\int d^3x \{ \frac{1}{2} B^2+ 
\frac{1}{2}Tr\{(D_i\phi_1)^2\}+  \frac{1}{2}Tr\{(D_i\phi_2)^2\} \} \quad.
\end{equation}
To be able to get the BPS equations we first make the following
rescalings: $\phi_1\to \cosh{u} ~\phi_1$,$\phi_2\to \sinh{u} ~\phi_2$,
$x_i \to \frac{1}{\sqrt{\cosh{2u}}}~x_i$ and
$A_i\to\sqrt{\cosh{2u}}~A_i$, where
$Tr\{\phi_2(\infty)^2\}=\sinh^2{u}$. Note the similarity with the
usual BPS dyon. Now we can write the energy in the following form:
\begin{equation}
E=\frac{1}{\sqrt{\cosh{2u}}}\int d^3x \{\cosh^2{u}~( \frac{1}{2} B^2+
\frac{1}{2} Tr\{(D_i\phi_1)^2\}) +\sinh^2{u} ~( \frac{1}{2} B^2+
\frac{1}{2} Tr\{(D_i\phi_2)^2\})\}\quad.
\end{equation}
From this we get the usual BPS equations for the monopole twice, one
for $\phi_1$ and one for $\phi_2$. They reduce to one set of BPS
equations under the assumption $\phi_1=\phi_2$\footnote{This is
possible since the rescaled fields $\phi_1$ and $\phi_2$ have the same
boundary conditions.}, whose solution is well known. The energy of the
monopole is simply given by:
\begin{equation}
E_{mon}=\sqrt{\cosh{2 u}}~ E_{mon;~u=0}\quad,
\end{equation}
where $E_{mon;~u=0}$ is the energy of the monopole in absence of
$\phi_2$.
\\[2mm]
This shows that the core structure of the monopole is effected by the
boundary conditions of the $\phi_2$ field and that there is no
DVS. This in contrast with the results found in one and two
dimensions.

\section{Conclusion and outlook}
In this paper we showed the possibility of dynamical vacuum selection
(DVS) in one dimensional field theories with flat directions. We
examined this DVS with the help of a specific model. For this model we
proved that there is only one specific boundary condition which allows
a static kink solution in an infinite space. In a finite space any
boundary condition allows the formation of a static kink. With the
help of a relaxation program we numerically determined these
restricted kinks. They can be very well described by one specific kink
with a scalar cloud on each side of the kink. This description of the
restricted kink also correctly predicts the position of the kink with
respect to the boundaries of the space as a function of the boundary
conditions. Using a numerical simulation we examined the dynamical
properties of configurations with boundary conditions which do not
allow a static solution in an infinite space. These simulations
confirm the DVS, which was expected to occur, from the result of the
restricted kinks and the field equation for the modulus field.

It should be clear that DVS in one dimension is not specific for the
one dimensional model we considered. It is a general feature of one
dimensional models with flat directions. The argument relies crucially
on the the scaling of the energy of the tails, the scalar clouds, of
the restricted kinks. In these scalar clouds only the modulus field
changes. More generally this shows that DVS is only possible in one
and, as was shown before \cite{penin}, in two dimensions, but not in
three or higher dimensions. For completeness we briefly mentioned the
two dimensional case and included an explicit example of a three
dimensional model where DVS does not work.

It remains of course possible that a model has more then one possible
static configuration in an infinite space \cite{pogosian}, but the DVS
will just pick out one of the possible static solitons and including
tunneling it will eventually always select the vacuum corresponding to
the static soliton with the lowest energy. Although this DVS selects a
lowest energy static soliton solution there can still be a vacuum and
core degeneracy left if the groundstate of the topologically non
trivial sector is degenerate, as was for example found in
\cite{achucarro}.

It would be interesting to study DVS, due to the presence of a
soliton, in a theory where the selected vacuum is lifted from the
classical vacuum manifold by quantum mechanical (or thermal)
corrections. Obviously this cannot happen if a supersymmetric BPS
soliton is selected by the DVS, since the energy of such a BPS soliton
is protected against quantum mechanical corrections. On the length
scale where the quantum mechanical corrections to the vacuum manifold
would become important the DVS alters and most likely a type of
restricted soliton will become the new selected vacuum. The tail(s) of
these restricted solitons will presumably have a length of the order
determined by the quantum mechanical corrections to the classical
vacuum manifold.\\[2mm] We would like to thank A. Ach\'ucarro,
L. Pogosian and T. Vachaspati for discussions. One of us (J.S.)
thanks the ESF, COSLAB program, for supporting the participation in
the COSLAB 2002 workshop.

\end{document}